\documentclass[12pt]{article}
\usepackage{amsmath, latexsym, amssymb, amscd}
\numberwithin{equation}{section}

\newcommand{\R}{\mathbb R}
\newcommand{\C}{\mathbb C}

\newcommand{\p}{\partial}

\newcommand{\Bold}[1]{{\boldsymbol{\mathit{#1}}}}

\newcommand{\M}{\mathbb{M}}

\newcommand{\QED}{\hspace{.2in}\square\newline}

\newtheorem{example}{Example}[section]
\newtheorem{theorem}{Proposition}[section]

\newtheorem{definition}{Definition}[section]

\newtheorem{thesis}{Thesis}[section]

\begin{document}

\title{Apparent Geometry from the\\ Quantum Mechanics of $Sp(8,\mathbb{C})$}
\author{J. LaChapelle}

\maketitle

\begin{abstract}
Restricting attention to kinematics, we develop the $C^\ast$-algebraic quantum mechanics of $Sp(8,\mathbb{C})$. The non-compact group does double duty: it furnishes the quantum Hilbert space through induced representations, and it spawns the quantum $C^\ast$-algebra through a crossed product construction. The crossed product contains operators associated with the lie algebra of $Sp(8,\mathbb{C})$ whose spectra can be interpreted as a $\mathrm{dim}_{\mathbb{C}}=20$ non-commutative phase space with a dynamical, commutative $\mathrm{dim}_{\mathbb{C}}=10$ configuration subspace and an internal $U(4,\mathbb{C})$ symmetry. The construction realizes quantization without first passing through the classical domain, and it exhibits apparent geometry.
\end{abstract}

\section{Introduction}
We want to draw a distinction between `emergent' geometry/gravity and what we will call `apparent' geometry. The former has been proposed and widely studied in several incarnations; string theory, AdS/CFT, matrix models, entanglement/entropy, and non-commutative/fuzzy geometry. The existing literature is extensive: for a representative sample see \cite{BFFS}--\cite{STY}. These approaches have at least two things in common. First, (to the best of our knowledge) they all start with an underlying dynamical classical model and then quantize according to various well-established methods. Second, the geometry/gravity is a consequence of the dynamics. The term `emergent', then, conveys the notion that geometry and/or gravity are in some sense the classical approximation to certain quantized dynamics. The trick, of course, is to find the proper underlying classical model --- including dynamics --- and subsequent quantization mapping.

It is sensible to wonder if one can construct realistic quantum dynamics without appealing to an underlying classical model: let's call it direct quantization. Our aim here is to realize direct quantization based on two well-established principal pillars; algebraic quantum mechanics and symmetry. Algebraic quantum mechanics dictates a $C^\ast$-algebra with states and an associated Hilbert space. Symmetry describes discernable patterns in the composition of observations/measurements. To link the two, we assume: (i) a closed quantum system is governed by an abstract unital $C^\ast$-algebra $\mathfrak{A}_L$ whose group of units $G$ characterizes observed/measured system symmetries, and (ii) quantum dynamics is effected by inner automorphisms of $\mathfrak{A}_L$ generated by $G$.

We plan to implement direct quantization by constructing a crossed product \cite{W} based on $G$ and its associated representation-furnishing Hilbert space $\mathcal{H}$. The crossed product is presumed to be a concrete realization of the abstract algebra $\mathfrak{A}_L$. Similar to non-relativistic quantum mechanics based on the Heisenberg group, we can then identify certain operators coming from our choice of $G$ that, loosely speaking, behave like configuration and momentum variables. But unlike Heisenberg, momentum-type operators associated with $G$ combine non-trivially to yield expectation values that generate time-dependent symmetric forms, anti-symmetric forms, and almost complex structures on configuration spectra of a suitable set of operators with respect to a suitable reference state. These objects can be employed to construct metrics and symplectic forms on pertinent topological spaces of spectra. In this sense, the geometry `appears' via observation/expectation values. It doesn't emerge; as it relies only on the \emph{kinematics} of the direct quantization. (The assumption of inner automorphism-inducing dynamics then leads to evolving geometry which connotes gravity. However, in this paper we confine attention to the kinematics only, so the possible link between evolving geometry and Einstein gravity is not addressed.)

A brief overview of the paper follows: We begin by describing our proposal for direct quantization in \S2. The abstract algebra $\mathfrak{A}_L$ supplies a topological group of units $G$ that induces a countable family $G_\Lambda=\{G_\lambda,\lambda\in\Lambda\}$ of locally compact topological groups on which the Born rule can be implemented. In a sense, one can imagine $G_\Lambda$ as representing certain measurable/observable elements in $G$. The physically relevant representations of $G_\Lambda$ allow to construct a Hilbert space $\mathcal{H}$ and its associated algebra of linear, bounded operators $L_B(\mathcal{H})$. Choosing $G_\Lambda=Sp(8,\C)$, which is a non-compact group, we use the method of induced representations in \S3 to realize $\mathcal{H}$ and $L_B(\mathcal{H})$ in a fiber bundle framework and give a coherent state interpretation of states $\psi\in\mathcal{H}$ in \S4. With these preliminaries in place, \S5 constructs the crossed product $L_B(\mathcal{H})\rtimes Sp(8,\C)$ which is presumed to model the abstract quantum algebra $\mathfrak{A}_L$. Since the algebra is supposed to govern a closed quantum system, we also hypothesize that dynamics are induced by inner automorphisms. Finally, in \S6 we show how the Lie algebra of $Sp(8,\C)$ is associated with operators in the crossed product that, via the Born rule, lead to the appearance of physically realistic geometry.

We will not go into the original motivation that led to the identification of $Sp(8,\C)$ as a prospective symmetry group. But remark that $Sp(2n,\R)$ has been frequently investigated in the literature for obvious reasons; e.g. see \cite{RW,ADMS,WU}. The direct quantization presented in this paper is an abridged version of \cite{LA4,LA2,LA1} where the duties of crossed products are assumed by functional Mellin transforms. A more thorough treatment of the quantum mechanics of $Sp(8,\C)$ based on functional Mellin transforms is given in \cite{LA5}.  Crossed products have been used by \cite{LANDS1,LANDS2} to quantize homogenous spaces. The ideas in \cite{AG,KapW,BDH} (and to a lesser extent \cite{WHEEL,MMP}) echo the direct quantization and `apparent' geometry notions in this paper.

\section{Direct quantization}\label{preliminaries}
Suppose an abstract unital $C^\ast$-algebra $\mathfrak{A}_L$ equipped with a Lie bracket structure governs some  closed quantum system. A particularly interesting subset of elements of $\mathfrak{A}_L$ is its set of units.  Let $A_L$ be a (sub)group of units of $\mathfrak{A}_L$ and let $G$ denote a topological group isomorphic to $A_L$. By definition, $G$ is a topological linear Lie group since $A_L$ is endowed with a Lie bracket.\cite[def. 5.32]{HM} Evidently, if we can identify $G$ through observation, then we have a good start on finding $\mathfrak{A}_L$.

But in general $G$ is not locally compact, so there is no associated Haar measure with which to extract representable or measurable objects. We require, then, some rationale to obtain locally compact topological groups from $G$ --- the Born rule: We assume observation corresponds to measurable elements in $\mathfrak{A}_L$ and, for suitable elements, a probability interpretation can be assigned to the measurements. To make this more explicit, let $G_{\Lambda}:=\{G_{\lambda},\lambda\in\Lambda\}$ represent a \emph{countable} family of locally compact topological Lie groups $G_{\lambda}$ indexed by group epimorphisms $\lambda:G\rightarrow G_{\lambda}$. Remark that $G_{\lambda_i}$ may or may not be isomorphic to $G_{\lambda_j}$, and since $\lambda$ is surjective, one can view $G_{\lambda}$ as a subset $G_{\lambda}\subset G$. Hence implementing the Born rule in this context represents a `topological localization'.\footnote{We will be purposely nonspecific about the set $\Lambda$, because it depends on the particular quantum system under consideration. But in general it represents constraints, state preparation/observation, or any other system particulars that one must specify to implement the Born rule.}

\begin{example}
 An elementary illustration of `topological localization' is the familiar Feynman path integral for paths in $\R^n$. Here $G$ is the group, under point-wise addition, of Gaussian\footnote{By Gaussian paths we mean the pointed paths are characterized by a mean and covariance.} pointed paths $X_a\ni (x,t_a)\rightarrow(\C^n,x_a)$ where $t_a\in\R$, $x(t_a)=x_a\in \R^n$. $X_a$ is an infinite-dimensional abelian topological group when endowed with a suitable topology. The corresponding path integral over $X_a$ is a formal object. But as soon as one imposes a constraint on the loose ends of the paths, for example $x(t_b)=x_b\in\R^n$ which `pins' them to a single point, the group `localizes' to a finite-dimensional group $X_{a,b}$.\footnote{To see this, parametrize the space of Gaussian pointed paths by mean and covariance. Fixing the loose end-point fixes the mean, and the covariance is then parametrized by points in $\R^n$. Consequently, the moduli space of Gaussian pointed paths with both end-points fixed in $\R^n$ is congruent to $\R^n$.} Being a finite-dimensional topological vector space, it is automatically locally compact: The corresponding path integral can now be explicitly integrated. Physically, this corresponds to measurement of a transition amplitude.

 There are of course many other constraints that one can impose on a given system. These constitute the set $\Lambda$, and a particular choice of $\lambda\in\Lambda$ leads to a particular and explicit evaluation of the path integral over $X_a$.
\end{example}

The point is, $G$ inherits a Lie bracket structure from $\mathfrak{A}_L$ that can only be glimpsed as a member of $G_{\Lambda}$ through observation/measurement of a particular system. Given that $G_{\Lambda}$ has been discerned, the plan is to utilize representations of $G_{\Lambda}$ to generate a pertinent Hilbert space $\mathcal{H}$  and its associated $C^\ast$-algebra of linear, bounded operators $L_B(\mathcal{H})$. They can be realized explicitly through a fiber bundle formulation.

With these structures, we propose to model the abstract algebra $\mathfrak{A}_L$ by the crossed product $L_B(\mathcal{H})\rtimes G_{\Lambda}$.\footnote{Why the crossed product instead of $L_B(\mathcal{H})$? Because the latter includes only linear, bounded operators and we suspect evolution of a quantum system is controlled by a broader collection of operators.} Of course this crossed product is not likely to be equivalent to $\mathfrak{A}_L$. But in practice one  doesn't know $\mathfrak{A}_L$ explicitly anyway: The idea is, we know enough if we know $G_{\Lambda}$. This motivates
 \begin{thesis}\label{first assumption}
 The abstract $C^\ast$-algebra that characterizes a closed quantum system can be modeled by a
 crossed products $L_B(\mathcal{H})\rtimes G_{\Lambda}$ where $G_{\Lambda}$ is a family of locally compact topological linear Lie groups and the Hilbert space $\mathcal{H}$ furnishes suitable (sub)-representations of $G_{\Lambda}$. Expectations of observables in $L_B(\mathcal{H})\rtimes G_{\Lambda}$ are taken with respect to $\mathcal{H}$. This constitutes the kinematic input to a quantum theory.
 \end{thesis}

It is then both natural and economical to suppose that dynamics are modeled by \emph{inner} automorphisms of $\mathfrak{A}_L$. After all, the closed system presumably evolves independent of any external input. And, presumably, observing dynamics leads to the discernment of $G_\Lambda$ in the first place. This motivates
\begin{thesis}
The dynamics of a closed quantum system are generated by a dynamical group $G^D$ such that $G^D_\Lambda\subseteq G_{\Lambda}$, and evolution is governed by continuous, time-dependent unitary {inner} automorphisms.
\end{thesis}
Notice that the topology on $G$ is not fixed so \emph{continuous} evolution potentially can look very different depending on $G_\lambda\subset G$. Beyond this point, we will not develop or consider (with one exception) quantum dynamics, because apparent geometry is evident already at the kinematic level.

\section{Induced representations}
The first task of direct quantization is to determine pertinent representations of the non-compact group $Sp\,(8,\C)$.

$Sp\,(8,\C)$ is a rank-$4$ reductive complex Lie group of $\mathrm{dim}_\C(Sp\,(8,\C))=36$ with Lie algebra $\mathfrak{Sp}(8)$ whose triangular decomposition is
\begin{equation}
\mathfrak{Sp}(8)=\mathfrak{G}_-\oplus\mathfrak{G}_0\oplus\mathfrak{G}_+=:\mathfrak{G}
\end{equation}
where
\begin{eqnarray}
&&\left[\mathfrak{G}_0,\mathfrak{G}_0\right]=0\notag\\
&&\left[\mathfrak{G}_+,\mathfrak{G}_-\right]\subseteq\mathfrak{G}_0\notag\\
&&\left[\mathfrak{G}_\pm,\mathfrak{G}_0\oplus\mathfrak{G}_\pm\right]
\subseteq\mathfrak{G}_\pm\;.
\end{eqnarray}
To render these brackets more explicit, let $S_{F}$ denote a Fock
space of \emph{bosonic} excitations above some vacuum. Define creation and
annihilation operators acting on this space by
\begin{equation}
c_\alpha c^\dag_\beta- c^\dag_\beta c_\alpha=\delta_{\alpha\beta}
\;\;\;;\;\;\; c^\dag_\alpha c^\dag_\beta- c^\dag_\beta
c^\dag_\alpha=0\;\;\;;\;\;\; c_\alpha c_\beta- c_\beta c_\alpha=0
\end{equation}
where $\alpha,\beta\in\{\pm1,\ldots,\pm4\}$ and $\dag$ indicates the
conjugate operator. With these operators, a basis of
$\mathfrak{Sp}(8)$ can be realized as\cite{JQ}
\begin{equation}
c_{\alpha,\beta}:=\frac{1}{\sqrt{1+\delta_{\alpha,-\beta}}}
\left(c^\dag_\alpha c_\beta+(-1)^{\alpha-\beta}c^\dag_{-\beta}
c_{-\alpha}\right)
\end{equation}
and rearranged as
\begin{eqnarray}\label{rearranged generators}
&&\mathfrak{h}_a:=c_{a,a}=n_a-n_{-a}\notag\\
&&\mathfrak{e}_a:=c_{a,-a}=\sqrt{2}c_a^\dag c_{-a}\notag\\
&&\mathfrak{e}_{-a}:=c_{-a,a}=\mathfrak{e}_a^\dag\notag\\
&&\mathfrak{e}_{a,b}:=c_{a,(a-b)}\notag\\
&&\mathfrak{e}^\dag_{b,\,a}=\mathfrak{e}_{a,b}=(-1)^{a+b}c_{(b-a),-a}
\end{eqnarray}
where $n_a:=c^\dag_a c_a$ and the indices $a,b\in\{1,\cdots,4\}$ with $a\neq b$. This arrangement characterizes the Borel subgroup and its induced
coset space with associated subalgebras
$\mathfrak{G}_0\cong\mathrm{span}_\C\{\mathfrak{h}_a\}$,
$\mathfrak{G}_+\cong\mathrm{span}_\C\{\mathfrak{e}_a,\mathfrak{e}_{a,b}\}$, and
$\mathfrak{G}_-\cong\mathfrak{G}_+^\dag$.

The Borel decomposition can be used to build up a Fock space of states giving rise to infinite-dimensional irreducible discrete series representations of $Sp(8,\C)$ (see e.g. \cite{HC}--\cite{KP}). The states associated with these irreducible representations clearly represent excitations due to the action of $\mathfrak{Sp}(8)$, but the physical interpretation of the excitations and their associated quantum numbers is not particularly evident.

Instead, to facilitate physical interpretation, we will choose a parabolic decomposition motivated by the fact that $\mathfrak{U}(4)$ is the maximal compact subalgebra of $\mathfrak{Sp}(8)$ (over $\R$) and the observation that there are ten mutually
commuting generators contained in
$\mathfrak{G}_-\oplus\mathfrak{G}_+$.

Consider
\begin{eqnarray}
&&\{\mathfrak{u}_{ab}\}
:=\{\mathfrak{h}_a,\mathfrak{e}_{a,-b},\mathfrak{e}^\dag_{a,-b}\}
,\;\;\;\;\;1\leq a< b\leq 4\notag\\
&&\{\mathfrak{\mathfrak{e}}_{ab},\mathfrak{\mathfrak{e}}_{ab}^\dag\}:=\left\{(\mathfrak{e}_{a},\mathfrak{e}_{a,+b}),
(\mathfrak{e}^\dag_{a},\mathfrak{e}^\dag_{a,+b}) \right\} ,\;\;\;\;\;1\leq a< b\leq
4\;.
\end{eqnarray}
The first set generates $U(4)$, and the set of generators
$\{\mathfrak{e}_{ab}\}$ (resp.$\{\mathfrak{e}_{ab}^\dag\}$)  mutually commute. They satisfy the commutation relations
\begin{eqnarray}\label{commutation relations}
&&[\mathfrak{e}_{ab}\,,\,{\mathfrak{e}}_{cd}^\dag]
=\delta_{ac}\mathfrak{u}_{db}+\delta_{ad}\mathfrak{u}_{cb}
+\delta_{bc}\mathfrak{u}_{da}+\delta_{bd}\mathfrak{u}_{ca}\notag\\
&&[\mathfrak{u}_{ab}\,,\,{\mathfrak{u}}_{cd}]
=\delta_{bc}\mathfrak{u}_{ad}-\delta_{ad}\mathfrak{u}_{cb}\notag\\
&&[\mathfrak{u}_{ab}\,,\,{\mathfrak{e}}_{cd}]
=\delta_{bc}\mathfrak{e}_{ad}+\delta_{bd}\mathfrak{e}_{ac}\notag\\
&&[\mathfrak{u}_{ab}\,,\,{\mathfrak{e}}_{cd}^\dag]
=-\delta_{ac}{\mathfrak{e}}_{bd}^\dag
-\delta_{ad}{\mathfrak{e}}_{bc}^\dag\notag\\
&&[\mathfrak{e}_{ab}\,,\,{\mathfrak{e}}_{cd}]=
[{\mathfrak{e}}_{ab}^\dag\,,\,{\mathfrak{e}}_{cd}^\dag]=0\;.
\end{eqnarray}

Let $\varrho':\mathfrak{Sp}(8)\rightarrow L(\mathcal{V})$ be a representation
with $\mathcal{V}$ a $\mathfrak{G}$-module. The
triangular decomposition of the algebra induces a decomposition of
$\mathcal{V}$ by
\begin{equation}\label{weight decomposition}
\mathcal{V}=\bigoplus_{w} \mathcal{V}_{(w)}
\,,\;\;\;\mathcal{V}_{(w)}:=\{\Bold{v}\in \mathcal{V}
:\varrho'(\mathfrak{h}_a)\Bold{v}=w_a\Bold{v}\}\,,\;\;\;a\in\{1,\ldots,4\}
\end{equation}
where $\mathfrak{h}_a\in\mathfrak{G}_0$ and ${w}=\{w_1,\ldots,w_4\}$
is a weight in the basis of fundamental weights
composed of complex eigenvalues $w_a\in\C$. It is well known that a particular $\mathcal{V}$ can be generated by acting with raising operators $\mathfrak{g}_ +\in\mathfrak{G}_+$ on a dominant-integral lowest-weight vector $\Bold{v}_{w_-}$.\cite{FU} Call this vector space $\mathcal{V}_{{w_-}}$.

Now, there is a distinguished subalgebra of $\mathfrak{Sp}(8)$; its
maximal compact subalgebra $\mathfrak{U}(4)$. Let
$\mathcal{V}_{({\mu})}\subset\mathcal{V}_{{w_-}}$ denote the submodule
generated by $\mathfrak{U}(4)$ acting on the dominant-integral lowest-weight vector
$\Bold{v}_{w_-}$. The submodule $\mathcal{V}_{({\mu})}$ then furnishes an
irreducible representation (IR) $\bar{\varrho}'$ of $U(4)$ where
${\mu}=[\mu_1,\ldots,\mu_4]$ is a partition based on $w_-$ that labels
the representation and $\bar{\varrho}'$ is a restricted representation of $\varrho'$. Since $w_-$ is a lowest weight,
$\mathcal{V}_{({\mu})}$ is an invariant sub-space with respect to the
subalgebra $\mathfrak{P}:=\mathfrak{G}_-\cup\mathfrak{U}(4)$, that is
${\bar{\varrho}}'(\mathfrak{P})\mathcal{V}_{({\mu})}\subseteq\mathcal{V}_{({\mu})}$.
From this, one obtains representations of $\mathfrak{P}$ based on lowest-weight IRs of $U(4)$  that are labeled by partitions $[\mu_1,\ldots,\mu_4]$. Recall that $U(4)$ enjoys \emph{both} boson and fermion representations.(e.g. \cite[pg. 500]{G})

At this point, we have representations of the subalgebra $\mathfrak{P}$ and a parabolic decomposition
\begin{equation}
\frac{\mathfrak{Sp}(8)}
{\mathfrak{P}}=:\frac{\mathfrak{Sp}(8)}
{\mathfrak{Z}_-\oplus\mathfrak{U}(4)}=:\mathfrak{Z}_+\;.
\end{equation}
From here, construct its associated complex coset space $Z:=Sp\,(8,\C)/P^C$ with $\mathrm{dim}_\C(Z)=10$.\footnote{Note that the elements of $\mathfrak{Z}_+$ mutually commute so it is natural to interpret $Z$ as parametrizing compatible quantum observables. We do not address dynamics here, but one can anticipate the elements of $\mathfrak{P}$ as the generators of `external' and `internal' dynamics relative to $Z$ based on the commutation relations (\ref{commutation relations}).}
The space $Z$ furnishes a convenient means to construct induced URs.

The goal is to construct unitary representations (URs) of the simply connected group $Sp(8,\C)$. This is a non-compact group, and we can't simply exponentiate a representation of its algebra because relevant representations are generally infinite-dimensional in this case. The method used to construct URs relies on Mackey's theory of induced representations\cite{M1,M2,M3} which leverages $\bar{\varrho}'$. We give an outline of the steps:
\begin{description}
\item[step 1:]Find the \emph{basic} dominant-integral lowest-weight
modules of $\mathfrak{Sp}(8)$.  There are four: $\{\mathcal{V}^{(0)}_1,\mathcal{V}^{(1)}_8,\mathcal{V}^{(2)}_{27},\mathcal{V}^{(3)}_{48},\mathcal{V}^{(4)}_{42}\}$ where the subscript denotes the dimension of the module and the superscript labels the representation. The trivial representation $\mathcal{V}^{(0)}_1$ has been included in this list, because it will eventually represent the quantum vacuum. The defining module is $\mathcal{V}^{(1)}_8$, and the adjoint module is $\mathcal{V}^{(0)}_1\oplus\mathcal{V}^{(1)}_8\oplus\mathcal{V}^{(2)}_{27}$. Whether there are other relevant representations based on $\mathcal{V}^{(3)}_{48}$ and $\mathcal{V}^{(4)}_{42}$ is unclear, but there is no reason not to expect them.

\item[step 2:]For each relevant representation, identify the dominant-integral lowest-weight vector  and generate the $\mathfrak{P}$ invariant sub-space $\mathcal{V}_{(\mu)}\subset\mathcal{V}_{w_-}$ for all relevant
unitary IRs of $U(4)$ by acting on the dominant-integral lowest-weight vector $\Bold{v}_{w_-}$. $U(4)$ being compact,
its unitary IRs have finite dimension, and, since they are dominant-integral, the various $\mathcal{V}_{(\mu)}$ posses a positive definite Hermitian inner product.

\item[step 3:]Recall that the action of $\mathfrak{P}$
leaves $\mathcal{V}_{(\mu)}$ invariant. So maximum efficiency
(associated with the impending induced representation) obtains
through the factorization $Sp\,(8,\C)/P^\C$ which utilizes the
finite-dimensional $\mathcal{V}_{(\mu)}$. Accordingly,  extend\footnote{The extension from $U(4)$ to $P$ is trivial because, as follows from the commutation relations, $\mathfrak{Z}_-$ annihilates the lowest weight element in $\mathcal{V}_{(\mu)}$. Subsequently, the representation on $P$ can always be extended to $P^\C$.} the relevant UIRs of $U(4)$ to $P^\C$. Since the ten elements
in the factor algebra $\mathfrak{Z}_+$ mutually
commute,  we can anticipate they will yield a basis for
compatible quantum observables.

\item[step 4:]Construct the principal coset bundle
$(\mathcal{P},Z_\p,\breve{pr},P^\C)$ and its associated vector bundle
$(\mathcal{V},Z_\p,pr,\mathcal{V}_{(\mu)},P^\C)$ where $\mathcal{P}\equiv Sp\,(8,\C)$ and the base space
may be a submanifold of the homogeneous coset space; $Z_\p\hookrightarrow Z:=Sp\,(8,\C)/P^\C$.
Recall that a point $g\in \mathcal{P}$ is an admissible map
$g:\mathcal{V}_{(\mu)}\rightarrow\mathcal{V}$. Since we are
stipulating unitary IRs of $U(4)$, there is a unique lowest weight
$\Bold{v}_{w_-}\in\mathcal{V}_{(\mu)}$ invariant under the right action of $P$ so that
$g(\Bold{v}_{w_-})$ can be identified with the zero-section in
$\mathcal{V}$. It is important that  the elements of
$\mathfrak{Z_+}$ mutually commute since then
$\mathrm{exp}\{\mathfrak{z}_+\}(\mathcal{V}_{(\mu)})$
induces a foliation of $\mathcal{V}$ compatible with the fiber
structure, i.e. leaves are homeomorphic to $Z$.

\item[step 5:]For each relevant $\mathcal{V}^{(r)}_{(\mu)}$ (that is, $\mathcal{V}_{(\mu)}$ labeled by the  representation $r$ which is determined by $U(4)$ quantum numbers and the right action of $\mathfrak{Z}_-$ on $Z$), consider the
 equivariant, continuous, compactly-supported vector-valued maps $\breve{\psi}\in
C_C(\mathcal{P},\mathcal{V}^{(r)}_{(\mu)})$ with finite norm
\begin{equation}\label{p-norm}
\|\breve{\psi}\|_{L^2} =\left(\mathrm{tr}\,\int_{Sp\,(8,\C)}
|\breve{\psi}(g)|^2\;d\mu\right)^{1/2}
\end{equation}
where $d\mu$ is a left Haar measure and the trace is with respect to the scalar product on
$\mathcal{V}^{(r)}_{(\mu)}$. The induced unitary representations are then defined by
\begin{equation}\label{induction}
\mathrm{UInd}_{P^\C}^{Sp\,(8,\C)^{(r)}}
=\{\breve{\psi}\in
L^2(\mathcal{P},\mathcal{V}^{(r)}_{(\mu)})\,|\,\breve{\psi}(g\,p)
=N(p)\bar{\varrho}(p^{-1})\breve{\psi}(g)\}
\end{equation}
where $p\in P^\C$, the normalization $N^2(p):={\triangle_{P}(p)/
\triangle_{Sp\,(8,\C)}(p)}$ with modular function
$\triangle_G(g)=|\mathrm{det}\,Ad_G(g)|$, and the continuous map $\bar{\varrho}:P^\C\rightarrow
L(\mathcal{V}^{(r)}_{(\mu)})$ is a unitary lowest-weight IR from step 3.

\item[step 6:]Construct the Whitney sum bundle
\begin{eqnarray}
\mathcal{W}_{\mathcal{V}}&:=&(\bigoplus_r\mathcal{V}^{(r)},Z_\p,pr,
\bigoplus_r\mathcal{V}^{(r)}_{(\mu)},P^\C)\notag\\
&=:&\left(\mathcal{W},Z_\p,pr, \mathcal{W}_{(\Bold{\mu})},P^\C\right)
\end{eqnarray}
using all relevant unitary IR
modules $\mathcal{V}^{(r)}_{(\mu)}$. The typical fiber $\mathcal{W}_{(\Bold{\mu})}$ is Hilbert.

The induced UR module is
\begin{equation}
\mathrm{UInd}_{P^\C}^{Sp\,(8,\C)}=\bigoplus_r\mathrm{UInd}_{P^\C}^{Sp\,(8,\C)^{(r)}}\;,
\end{equation}
and the induced UR $\rho:Sp\,(8,\C)\rightarrow
L_B(\mathrm{UInd}_{P^\C}^{Sp\,(8,\C)})$, which will \emph{not} be irreducible in general, can be expressed as
\begin{equation}
(\rho({g})\breve{\psi})(g_o)=\breve{\psi}(g^{-1}g_o)
=:\breve{\psi}_{{g}}(g_o)
\end{equation}
where $g_o,{g}\in Sp\,(8,\C)$.
\end{description}

\section{Hilbert space}
From the induced UR module $\mathrm{UInd}_{P^\C}^{Sp\,(8,\C)}$, we want to construct the kinematic quantum Hilbert space $\mathcal{H}$ of state vectors.

Since the generators associated with the homogenous space $Z$ mutually commute, they can be simultaneously diagonalized --- meaning state vectors of the eventual corresponding quantum system can be parametrized by the \emph{smooth manifold} $Z$. Also, we can transfer the furnishing space of the induced representation because $\breve{\psi}$ and $\psi\in\Gamma( Z,\mathcal{W})$ are related by $\breve{\psi}(g)=g^{-1}\circ\psi(z)$ with $\breve{pr}(g)=z=gz_0$
where $z_0$ is a choice of origin in $Z$. If a canonical local
section $\sigma_i$ on the principal bundle is chosen relative to a
local trivialization $\{U_i,\varphi_i\}$, then $\breve{\psi}$ and
$\psi$ are \emph{canonically} related, and we can identify $
\psi\equiv\sigma_i^\ast\breve{\psi}$.\cite{LA1,C-B}

Since $\breve{pr}(g\sigma_i(z))=g\breve{pr}(\sigma_i(z))=gz$, then $g\sigma_i(z)$
must be a point in the fiber over $gz$, i.e.
$g\sigma_i(z)=\sigma_i(gz)p$ for some $p\in P^\C$. Hence, using
canonical local sections relative to a given trivialization yields a
canonical induced representation on $\Gamma( Z,\mathcal{W})$;
\begin{eqnarray}
(\rho({g}){\psi})(z)&= &(\rho({g})\breve{\psi})(\sigma_i(z))\notag\\
&=&\breve{\psi}(g^{-1}\sigma_i(z))\notag\\
&=&N(p)\bar{\varrho}(p^{-1})\sigma_i^\ast\breve{\psi}(g^{-1}z)
=N(p)\bar{\varrho}(p^{-1})\psi(g^{-1}z)\notag\\
&&\hspace{1.55in}=:N(p)\bar{\varrho}(p^{-1})\psi_g(z)
\end{eqnarray}
where $p$ depends on $g$ and $z$ through $\sigma_i$.

Let $\mathcal{W}_z$ denote the fiber in $\mathcal{W}$ over the point $z\in Z_\p$, and let $(z,\Bold{v}_{w_{g_o}})$ denote  the representative of $\psi(z)$ in a local trivialization. The induced unitary representation $\rho$ defines a $\ast$-homomorphism $\bar{\pi}_z:\rho(G)\subset L_B(\mathrm{UInd}_{P^\C}^{Sp\,(8,\C)})\rightarrow L_B(\mathcal{W}_z)$ by
\begin{equation}
\left(\bar{\pi}_z(\rho(g))\right)\Bold{v}_{w_g}:=(\rho(g)\psi)(z)\;\;\forall g\in G
\end{equation}
This extends to ${\pi}_z:L_B(\mathrm{UInd}_{P^\C}^{Sp\,(8,\C)})\rightarrow L_B(\mathcal{W}_z)$ by the functional calculus (for suitable functions $t$)
\begin{equation}\label{pi}
(\pi_z (t(\rho(g)))\Bold{v}_{w_g}=:\left(\pi_z ({T}(g))\right)\Bold{v}_{w_g}\;\;\forall g\in G\;.
\end{equation}

There is ambiguity in this group action associated with $\bar{\varrho}(p)$: It can be interpreted either as an arbitrary choice of local section $\sigma_i$ or an arbitrary choice of basis on each fiber $\mathcal{W}_z$. Since a particular choice of either is not physically relevant, physical states (to be associated with elements of $\mathcal{H}$) should not depend on the choice. This implies that, if we want $\psi$ to represent a physical state, it should be equivariant under the right action of $P^\C$. But this is just the fiber bundle statement of gauge invariance. Conclude that a physical state is represented by an equivalence class $[\psi]$ with equivalence relation $\psi(z)\sim \psi(z p)$ for all $p\in P^\C$ and $z\in U_i$.\footnote{It is important to keep in mind that the equivalence relation is strictly imposed only in a local trivialization $\{U_i,\varphi_i\}$ since it springs from ambiguity of the choice of local section --- it is not meant to be a global statement.}

The condition that an entire equivalence class $[\psi]\in\mathcal{H}$ represents a physical state vector is readily handled in the bundle framework: just insist that the equivariant $\mathrm{p}$-forms  used to define the induced representation are \emph{horizontal} with respect to some chosen connection on $\mathcal{P}$. In effect this simply means that the kinematics obeyed by $\mathrm{p}$-forms  are defined in terms of an exterior covariant derivative $D$ on $\mathcal{V}$ associated with the choice of connection. The connection allows a consistent basis identification between $\mathcal{W}_z$ and $\mathcal{W}_{(\Bold{\mu})}$ so that $\pi_z$ determines ${\pi_{(\Bold{\mu})}}:L_B(\mathrm{UInd}_{P^\C}^{Sp\,(8,\C)})\rightarrow L_B(\mathcal{W}_{(\Bold{\mu})})$.

\begin{theorem}
Let $\mathcal{H}_{hor}\subseteq\mathcal{H}$ be the span of $\mathrm{span}_\C\{[\psi]\}$ for all $[\psi]\in\mathcal{H}$. Then $\mathcal{H}_{hor}$ is a Hilbert space.
\end{theorem}

\emph{Proof}:
The module
$\mathcal{H}=L^2(Z,\mathcal{W})\subset\Gamma(Z,\mathcal{W})$ furnishes a UR of
$Sp\,(8,\C)$ with sub-space $\mathcal{H}_{hor}\subseteq\mathcal{H}$ spanned by $[\psi]\equiv\psi_{hor}$. Use the quasi-invariant measure $\mu_{P^\C}$ on $Z$ and the Hermitian inner product on
$\mathcal{W}_{(\Bold{\mu})}$ to construct a bundle metric on $\mathcal{W}$.
Equip $\mathcal{H}$ with the Hermitian inner product induced from
$\mathcal{W}_z$ (equivalently from the Haar measure on $\mathcal{P}$) according to
\begin{equation}\label{inner product}
\langle\psi_1|\psi_2\rangle_{\mathcal{H}}:=\int_{Z_\p}(\psi_1(z)|\psi_2(z))_{\mathcal{W}_z}\;d\mu_{P^\C}(z)\;.
\end{equation}
Complete $\mathcal{H}$ with respect to the associated norm. Then $\mathcal{H}$ is Hilbert; and, since the connection is a linear map, $\mathcal{H}_{hor}$ is also Hilbert.
$\QED$

Ostensibly, $\mathcal{H}_{hor}$ can be identified with the kinematic quantum Hilbert
space. Unfortunately, the associated induced URs are not irreducible in general. This can be problematic for the interpretation of $[\psi]$ and implementation of the Born rule.

\subsection{Coherent states}
As remedy, we propose to model $[\psi]$ as a coherent state (CS). The structure of the coset space $Z:=Sp\,(8,\C)/P^\C$ immediately suggests defining Perelomov-type CS. Many useful details regarding these and other types of CS can be found in \cite{DQ,BR}.

Recall that $g\in G$ can be viewed as an admissible map $g:\mathcal{W}_{(\Bold{\mu})}\rightarrow\mathcal{W}$. Given an open region $U_i\in Z_\p$, a local
trivialization $\{U_i,\varphi_i\}$ of the Whitney sum bundle,  and a local chart
$\phi:U_i\rightarrow\C^{10}$; a point $w\in
\pi^{-1}(U_i)\subset\mathcal{W}$ can be represented on
$\C^{10}\times\mathcal{W}_{(\Bold{\mu})}$ as
\begin{equation}
|\phi(z);\Bold{\mu}):=\left(\exp\left\{\frac{1}{2}
\sum_{a,b}{z}^\ast_{ab}\,\mathfrak{e}_{ab}\right\}\right)| \Bold{\mu})
\end{equation}
where the point $z\in U_i$ has coordinates
$\phi(z)={z}^\ast_{ab}\in \C^{10}$ with $a,b\in\{1,\ldots,4\}$ such that $a\neq b$, the vector $|\Bold{\mu})\in\mathcal{W}_{(\Bold{\mu})}$, and we used the coset decomposition to parametrize $g=\exp\{\frac{1}{2}
\sum_{a,b}{z}^\ast_{ab}\,\mathfrak{e}_{ab}\}\exp\{\mathfrak{p}^\C\}$.

To simplify notation a bit choose normal coordinates and write
$|\phi(z);\Bold{\mu})\rightarrow|z^\ast;\Bold{\mu})$. Then a physical state vector
$[\psi]\in\mathcal{H}$ can be modeled locally on
$U_i\times\mathcal{W}_{(\Bold{\mu})}$. Explicitly,
\begin{definition}
Given a local trivialization of the bundle
$\mathcal{W}_{\mathcal{M}}$, the CS model of a state vector
$[\psi]\in\mathcal{H}$ is defined by\footnote{This mixed bracket
notation is a bit strange: On one hand it should not be confused
with the Hilbert space inner product
$\langle\cdot\,|\cdot\,\rangle_{\mathcal{H}}$ or the inner product
$(\cdot\,|\cdot\,)_{\mathcal{W}_{(\Bold{\mu})}}$. On the other hand, it
emphasizes that the object it defines is a CS model of a
state vector. It must be kept in mind that the notation
$(\cdot\,|\cdot\,\rangle$ implicitly assumes a local trivialization,
and can be interpreted as the expression of a state vector in the
``$z\otimes \Bold{\mu}$ representation''.}
\begin{equation}
({z};\Bold{\mu}|\psi\rangle=:\Bold{\psi}_{\Bold{\mu}}(z) \equiv
\sigma_i^\ast\breve{\psi}(z)
\end{equation}
where $\sigma_i$ is the canonical local section and $z\in Z_\p\subseteq Z$. The space $Z_\p$ is determined by boundary conditions on $\Bold{\psi}_{\Bold{\mu}}(z)$.
\end{definition}
Similarly, the $\ast$-homomorphism defined in (\ref{pi}) has a CS realization:
\begin{definition}
The CS model of an operator ${O}\in L_B(\mathcal{H})$ is
defined by
\begin{equation}\label{CS operator}
({z};\Bold{\mu}|\,{O}\,\psi\rangle
=:\widehat{{O}}\,\Bold{\psi}_{\Bold{\mu}}(z)\;.
\end{equation}
\end{definition}

We call $\Bold{\psi}_{\Bold{\mu}}(z)$ a coherent state wave function
or coherent state for short. It is a column vector according to
the relevant UIRs of $U(4)$ collectively labeled by
$\Bold{\mu}=(\mu^{(r_1)},\ldots,\mu^{(r_n)})$. Since
$\mathcal{V}^{(r)}_{w_-}$ are unitary IRs,
$\Bold{\psi}_{\Bold{\mu}}(z)$ is comprised of components
$\Bold{\psi}_{\Bold{\mu}}(z)
=(\Bold{\psi}^{r_1}_{\mu}(z),\ldots,
\Bold{\psi}^{r_n}_{\mu}(z))$ that do not mix --- a kind-of
super selection. We will often restrict to a specific component and write
$\Bold{\psi}_{{\mu}}(z)=(z;\mu|\psi\rangle\in\mathcal{V}^{(r)}_{(\mu)}$ without indicating the representation $r$
for notational and conceptual simplicity.

\subsubsection{Matrix CS}
The isomorphism between the space of $z_{ab}$ parameters and the
vector space of complex symmetric $4\times 4$ matrices allows to write the
coherent state basis as
\begin{equation}
|z^\ast;\Bold{\mu})=|\Bold{Z}^\ast;\Bold{\mu})
:=\left(\exp\left\{\frac{1}{2}
\mathrm{tr}(\Bold{{Z}}^\ast\mathfrak{E}_+)\right\}\right)| \Bold{\mu})
\end{equation}
where $\Bold{{Z}}^\ast\in M_4^{sym}(\C)$ is comprised of the coordinates
${z}_{ab}$ and it is understood that $U_i$ is modeled on
$M_4^{sym}(\C)$ --- the space of symmetric $4\times 4$ matrices with complex components.

To implement this, define the symmetric matrices $\mathfrak{E}_+$
and $\mathfrak{E}_-$ with components  $\{\mathfrak{e}_{ab}\}$ and
$\{\mathfrak{e}^\dag_{ab}\}$ respectively, and $\mathfrak{E}_U$
comprised of $\{\mathfrak{u}_{ab}\}$. Form
the vector space $M_4^{sym}(\C)\otimes\mathcal{W}_{(\Bold{\mu})}$, and model $Z$ on
$M_4^{sym}(\C)$. Given a chart on $Z$ and a local trivialization on
$\mathcal{W}$, a point is represented by
\begin{equation}
\varphi_i(w)=|\Bold{Z}^\ast;\Bold{\mu}) =\left(\exp\left\{\frac{1}{2}
\mathrm{tr}(\Bold{{Z}}^\ast\mathfrak{E}_+)\right\}\right)|\Bold{\mu})\;.
\end{equation}
Now define;
\begin{definition}
A CS model of a state vector in the matrix picture is defined by
\begin{equation}
({\Bold{Z}};\Bold{\mu}|\psi\rangle=(\Bold{\mu}|
\left(e^{\frac{1}{2}
\mathrm{tr}\,({\Bold{{Z}}}\mathfrak{E}_-)}\right)\,
\psi\rangle=:\Bold{\psi}_{\Bold{\mu}}(\Bold{Z})\;,
\end{equation}
and the model of an operator (not necessarily bounded) is
\begin{equation}
({\Bold{Z}};\Bold{\mu}|{O}\,\psi\rangle
=:\widehat{{O}}\,\Bold{\psi}_{\Bold{\mu}}(\Bold{Z})\;.
\end{equation}
\end{definition}

Referring to \cite{DQ},\footnote{Note that our
notation differs a bit from \cite{DQ}: we put ${\mu}:={\lambda}+n/2$
where ${\lambda}$ is the lowest weight characterizing the $U(4)$
unitary IR and $n\geq 2d=8$.} an explicit CS realization of the Lie algebra
generators  in a local trivialization
$U_i\times\mathcal{W}_{(\Bold{\mu})}$ in the matrix picture  is given by
\begin{equation}\label{generators1}
\widehat{\mathfrak{E}}_-=\left(
\begin{array}{cccc}
2\,\frac{\p}{\p z^{1}} & \frac{\p}{\p z^{12}} & \frac{\p}{\p z^{13}}
& \frac{\p}{\p z^{14}} \\ \\
\frac{\p}{\p z^{12}} & 2\,\frac{\p}{\p z^{2}} & \frac{\p}{\p z^{23}
}& \frac{\p}{\p z^{24}} \\ \\
 \frac{\p}{\p z^{13}} & \frac{\p}{\p z^{23}} & 2\,\frac{\p}{\p z^{3}} & \frac{\p}{\p z^{34}}
 \\ \\
 \frac{\p}{\p z^{14}} & \frac{\p}{\p z^{24}} &\frac{\p}{\p z^{34}} & 2\,\frac{\p}{\p z^{4}} \\
  \end{array}
  \right)\otimes\Bold{I}=:\partial_{\Bold{Z}}\otimes\Bold{I}
\;,
\end{equation}
\begin{eqnarray}
\widehat{\mathfrak{E}}_+&=&\left[\Bold{Z}\partial_{\Bold{Z}}-(d+1)\right]\Bold{Z}
\otimes\Bold{I}+\Bold{Z}\otimes
\Bold{U}+(\Bold{Z}^{\mathrm{T}}\otimes
\Bold{U})^{\mathrm{T}}\notag\\
&=&\left[\Bold{Z}\partial_{\Bold{Z}}-(d+1)\right]\Bold{Z}
\otimes\Bold{I}+\Bold{Z}\otimes \Bold{U}+(\Bold{Z}\otimes
\Bold{U})^{\mathrm{T}}\notag\\
&=:&\left[\Bold{Z}\partial_{\Bold{Z}}-(d+1)\right]\Bold{Z}
\otimes\Bold{I}+Sym(\Bold{Z}\otimes\Bold{U})\;,
\end{eqnarray}
and
\begin{equation}\label{generators3}
\widehat{\mathfrak{E}}_U=\Bold{Z}\partial_{\Bold{Z}}\otimes\Bold{I}
+\Bold{I}\otimes\Bold{U}
\end{equation}
where
\begin{equation}
(\Bold{U})_{\Bold{\mu}'\,\Bold{\mu}}:=(\Bold{\mu}'|\varrho'(\mathfrak{E}_{\,U})|
\Bold{\mu})_{\mathcal{W}_{(\Bold{\mu})}}
\end{equation}
with $\mathfrak{E}_{\,U}$ the generators of $U(4)$. These Lie algebra generators do not belong to $L_B(\mathcal{H})$ but, restricted to a suitable domain, their unitary exponentials do.

In words, the
set of matrix-valued operators
$\{\widehat{\mathfrak{E}}_+,\widehat{\mathfrak{E}}_-,
\widehat{\mathfrak{E}}_U\}$ is a Perelomov-type CS model of the Lie algebra generators
$\{\mathfrak{e}_{ab},\mathfrak{e}^\dag_{ab},\mathfrak{u}_{ab}\}$ in the matrix
picture, and unitary exponentiation realizes an induced UR of $Sp\,(8,\C)$.

The \emph{non-trivial} reproducing kernel for $(z',z)\in U_i$
has been calculated explicitly\cite{DQ};
\begin{eqnarray}\label{overlap}
({\Bold{Z}'};\Bold{\mu}'|\Bold{Z}^\ast;\Bold{\mu})
=(\Bold{\mu}'|\rho\left(e^{\mathrm{tr}\Bold{B}\mathfrak{E}_U}\right)|\Bold{\mu})_{\mathcal{W}_{(\Bold{\mu})}}
&=:&(\Bold{K}(\Bold{Z}',\Bold{Z}^\ast))_{\Bold{\mu}'\,\Bold{\mu}}
\end{eqnarray}
where $e^{-\Bold{B}}=(\Bold{I}-\Bold{Z}'\Bold{Z}^\ast)$. The associated
resolution of the identity is
\begin{eqnarray}
{{Id}}
&=&\int_{U_i}|\Bold{Z}^\ast;\Bold{\mu})\; \mathbf{d\,\Bold{\sigma}}(z)\;(
{\Bold{Z}};\Bold{\mu}|
\end{eqnarray}
where\footnote{Clearly there are subtleties associated with the determinant term in $\Bold{P}(\Bold{Z})$, but we will ignore them here.}
\begin{equation}
\mathbf{d\,\Bold{\sigma}}(z)
:=\mathcal{N}\,\frac{\Bold{K}^{-1}(\Bold{Z},\Bold{Z}^\ast)}
{\mathrm{det}(\Bold{I}-{\Bold{Z}}\Bold{Z}^\ast)^{(4+1)}}
\;\;d\mu_{P^\C}(z)=:\Bold{P}(\Bold{Z})\;d\mu_{P^\C}(z)\;.
\end{equation}
$\mathcal{N}$ is a normalization constant and
$\widehat{{Id}}\,\Bold{\psi}_{\Bold{\mu}}(\Bold{Z})=({Id}\,\psi)(\Bold{Z})$
with ${Id}$ the identity operator on $\mathcal{H}$.

From these, one obtains the local CS superposition on $\mathcal{W}$;
\begin{eqnarray}
|\psi\rangle_i
&=&\int_{U_i}|\Bold{Z}^\ast;\Bold{\mu})\; \mathbf{d\,\Bold{\sigma}}(z)\;(
{\Bold{Z}};\Bold{\mu}|\psi\rangle
\end{eqnarray}
which must then be extended globally to $Z_\p\backslash S^n$ with $S^n$ the unit sphere in $Z$ and Dirichlet/Neumann boundary conditions on the sphere $({\Bold{Z}};\Bold{\mu}|\psi\rangle|_{S^n}=\mathit{\Psi}_{\Bold{\mu}}$. Similarly, assuming $\mathit{\Psi}_{\Bold{\mu}}=0$ for simplicity,
\begin{eqnarray}\label{operator symbol}
\langle\psi|{O}\,\psi\rangle_{\mathcal{H}}
&=&\int_{Z_\p\backslash S^n}\Bold{\psi}^\dag_{\Bold{\mu}'}(\Bold{Z}')\Bold{P}(\Bold{Z}')\,
\widehat{{O}}\,\Bold{\psi}_{\Bold{\mu}'}(\Bold{Z}')\;d\mu_{P^\C}(z')\notag\\
&=&\int_{Z_\p\backslash S^n}\int_{Z_\p\backslash S^n}\Bold{\psi}^\dag_{\Bold{\mu}'}(\Bold{Z}')\Bold{P}(\Bold{Z}')\,
(\Bold{Z}';\Bold{\mu}'|O|\Bold{Z}^\ast;\Bold{\mu})\,\Bold{P}(\Bold{Z})\Bold{\psi}_{\Bold{\mu}}(\Bold{Z})\;
d\mu_{P^\C}(z,z')\,\notag\\
&=:&\int_{Z_\p\backslash S^n}\int_{Z_\p\backslash S^n}\Bold{\psi}^\dag_{\Bold{\mu}'}(\Bold{Z}')\,
\Bold{K}_{O}(\Bold{Z}',\Bold{Z}^\ast)\,\Bold{\psi}_{\Bold{\mu}}(\Bold{Z})\;
d\mu_{P^\C}(z,z')\;.
\end{eqnarray}
Note that expectations $\langle\cdot|\cdot\rangle_{\mathcal{H}}$ depend implicitly on $Z_\p$. Therefore, they can be referred to as global or as local a system as desired provided one knows $\psi|_{Z_\p}$.

\subsubsection{CS vacuum}

Let $\Bold{w}_-:=(w_-^{(r_1)},\ldots,w_-^{(r_n)})$
denote the collection of dominant-integral lowest weights. We define the ground state
by
$\breve{\psi}_0(g):=\Bold{v}_{\Bold{w}_-}\in\mathcal{W}_{(\Bold{\mu})}
\;\forall g\in Sp(8,\C)$. And we define a vacuum-state to be the ground state of a UR $\rho$ induced from the trivial partition $\Bold{\mu}=[\mu,\mu,\mu,\mu]$. In this case, $\mathcal{W}_{(\Bold{\mu})}\supset\mathcal{V}_1^{(0)}$ is irreducible and one-dimensional such that $(\rho(g)\breve{\psi}_0)(g)\propto\Bold{v}_{{\mu}}$ for all $g\in Sp(8,\C)$.

According to the definition, the CS model of the algebra generators acting on a ground state $\psi_0$ give $\widehat{\mathfrak{E}}_-\Bold{\psi}_{0}=0$, $\widehat{\mathfrak{E}}_+\Bold{\psi}_{0}=\Bold{Z}\otimes(\Bold{I}+Sym(\Bold{U}))\Bold{\psi}_{0}$, and $\widehat{\mathfrak{E}}_U\Bold{\psi}_{0}=\Bold{I}\otimes\Bold{U}\Bold{\psi}_{0}$. Obviously the vacuum-state is invariant under $\widehat{\mathfrak{E}}_-$ and $\widehat{\mathfrak{E}}_U$; as are $U(4)$ invariant ground states if they exist. This suggests a natural definition of the CS model of a
vacuum state vector;
\begin{definition}
The CS model of a vacuum-state vector $\varphi_0\in\mathcal{H}$ is
defined by
\begin{equation}
({z};\Bold{\mu}|\varphi_0\rangle=:\Bold{v}_{{\mu}}(z)
\equiv\Bold{v}_{\mu}\;\;\;\;\;\forall z\in Z\;
\end{equation}
such that $\langle\varphi_0|\varphi_0\rangle_{\mathcal{H}}=|\Bold{v}_{\mu}|$.
\end{definition}

\section{The algebra $\mathfrak{A}_L$}
The plan is to represent $\mathfrak{A}_L$ --- or rather its `shadow' under the Born rule --- by means of a crossed product.

The elements necessary to define a crossed product\cite{W} are: i) a dynamical system $(A,G,\varepsilon)$ where $A$ is a $C^\ast$-algebra, $G$ is a locally compact group, and $\varepsilon:G\rightarrow Aut(A)$ is a continuous homomorphism; ii) some Hilbert space $\mathrm{H}$; iii) an algebra representation $\pi:A\rightarrow L_B(\mathrm{H})$; and iv) a unitary, group representation $U:G\rightarrow U(\mathrm{H})$. The two representations are required to satisfy the `covariance condition'
\begin{equation}\label{covariance condition}
\pi(\varepsilon_g(a))=U(g)\pi(a)U(g)^\ast\;,\;\;\;\;g\in G\;,\;\;a\in A\;.
\end{equation}
With these elements, a $\ast$-representation of $C_c(G,A)$ (continuous compact morphisms $\mathrm{f}:G\rightarrow A$) on the Hilbert space $\mathrm{H}$ is supplied by the integral
\begin{equation}\label{crossed product}
\pi\rtimes U(\mathrm{f}):=\int_G\pi(\mathrm{f}(g))U(g)\;d\mu(g)
\end{equation}
where $\mathrm{f}\in C_c(G,A)$ and $\mu$ is a left Haar measure on $G$.

A product and involution are introduced on $C_c(G,A)$ according to
\begin{equation}
(\mathrm{f}_1\ast \mathrm{f}_2)(g):=\int_G \mathrm{f}_1(\tilde{g})\varepsilon_{\tilde{g}}(\mathrm{f}_2(\tilde{g}^{-1}g))\;d\mu(\tilde{g})
\end{equation}
and
\begin{equation}
\mathrm{f}^\ast(g):=\Delta(g^{-1})\varepsilon_g(\mathrm{f}(g^{-1})^\ast)
\end{equation}
where $\Delta$ is the modular function on $G$. Finally, completion of $C_c(G,A)$ with respect to the norm defined by
\begin{equation}
\|\mathrm{f}\|:=\mathrm{sup}\|\pi\rtimes U(\mathrm{f})\|
\end{equation}
is a $C^\ast$-algebra called the crossed product denoted by $A\rtimes_\varepsilon G$.\cite{W}

The crucial property of this construction is a one-to-one correspondence between non-degenerate covariant representations of $(A,G,\varepsilon)$  and non-degenerate representations of  $A\rtimes_\varepsilon G$ that preserve direct sums, irreducibility, and equivalence. So the $C^\ast$-algebra $A\rtimes_\varepsilon G$ can be used to model the quantum $C^\ast$-algebra encoded in the dynamical system $(A,G,\varepsilon)$ endowed with a covariant representation $(\pi,U)$.

In our case, we have $A\equiv L_B(\mathcal{H})$, $G\equiv Sp(8,\C)$, and we insist that $\varepsilon$ is an \emph{inner} automorphism. Further, we have $\mathrm{H}\equiv \mathcal{W}_{(\Bold{\mu})}$ so that $\pi\equiv\pi_{(\Bold{\mu})}$ and $U\equiv\pi_{(\Bold{\mu})}\cdot\rho=:\rho_{(\Bold{\mu})}$. Being $\varepsilon$ an inner automorphism, $(\pi_{(\Bold{\mu})},\rho_{(\Bold{\mu})})$ is clearly a covariant representation.
\begin{theorem}
Let $(L_B(\mathcal{H}),Sp(8,\C),\varepsilon)$ be a dynamical system with covariant representation $(\pi_{(\Bold{\mu})},\rho_{(\Bold{\mu})})$. Then $\pi_{(\Bold{\mu})}\rtimes\rho_{(\Bold{\mu})}(\cdot)$ is a $\ast$-representation of $L_B(\mathcal{H})\rtimes_\varepsilon Sp(8,\C)$ on $ \mathcal{W}_{(\Bold{\mu})}$.
\end{theorem}

\emph{Proof}:
Using the $\ast$-product and involution;
\begin{eqnarray}
\pi_{(\Bold{\mu})}\rtimes\rho_{(\Bold{\mu})}(\mathrm{f}_1\ast\mathrm{f}_2)
&=&\int_G\int_G\pi_{(\Bold{\mu})}(\mathrm{f}_1(g)\varepsilon_g(\mathrm{f}_2(g^{-1}\tilde{g}))
\rho_{(\Bold{\mu})}(\tilde{g})\;d\mu(g)\,d\mu(\tilde{g})\notag\\
&=&\int_G\int_G\pi_{(\Bold{\mu})}(\mathrm{f}_1(g)
\rho_{(\Bold{\mu})}(g)\pi_{(\Bold{\mu})}(\mathrm{f}_2(g^{-1}\tilde{g}))\rho_{(\Bold{\mu})}(g^{-1})
\rho_{(\Bold{\mu})}(\tilde{g})\;d\mu(g)\,d\mu(\tilde{g})\notag\\
&=&\int_G\int_G\pi_{(\Bold{\mu})}(\mathrm{f}_1(g)
\rho_{(\Bold{\mu})}(g)\pi_{(\Bold{\mu})}(\mathrm{f}_2(\tilde{g}))
\rho_{(\Bold{\mu})}(\tilde{g})\;d\mu(g)\,d\mu(\tilde{g})\notag\\
&=&\pi_{(\Bold{\mu})}\rtimes\rho_{(\Bold{\mu})}(\mathrm{f}_1)
\cdot\pi_{(\Bold{\mu})}\rtimes\rho_{(\Bold{\mu})}(\mathrm{f}_2)
\end{eqnarray}
and
\begin{eqnarray}
(\pi_{(\Bold{\mu})}\rtimes\rho_{(\Bold{\mu})}(\mathrm{f}))^\ast
&=&\int_G\left(\pi_{(\Bold{\mu})}(\mathrm{f}(g))\rho_{(\Bold{\mu})}(g)\right)^\ast\;d\mu(g)\notag\\
&=&\int_G\rho_{(\Bold{\mu})}(g)\pi_{(\Bold{\mu})}(\mathrm{f}(g^{-1})^\ast)\Delta(g^{-1})\;d\mu(g)\notag\\
&=&\int_G\pi_{(\Bold{\mu})}(\varepsilon_g(\mathrm{f}(g^{-1})^\ast\Delta(g^{-1}))\rho_{(\Bold{\mu})}(g)\;d\mu(g)\notag\\
&=&\int_G\pi_{(\Bold{\mu})}(\mathrm{f}^\ast(g))\rho_{(\Bold{\mu})}(g)\;d\mu(g)\notag\\
&=&\pi_{(\Bold{\mu})}\rtimes\rho_{(\Bold{\mu})}(\mathrm{f}^\ast)\;.
\end{eqnarray}
We won't require it here, but one can show $\|\pi_{(\Bold{\mu})}\rtimes\rho_{(\Bold{\mu})}(\mathrm{f})\|\leq\|\mathrm{f}\|_1$ and $\pi_{(\Bold{\mu})}\rtimes\rho_{(\Bold{\mu})}$ is non-degenerate if $\pi_{(\Bold{\mu})}$ is non-degenerate, i.e. if $\{\pi_{(\Bold{\mu})}(\mathrm{f})\Bold{v}:\,\Bold{v}\in\mathcal{W}_{(\Bold{v})}\}$ is dense in $\mathcal{W}_{(\Bold{\mu})}$.\cite[pg. 49]{W}
$\QED$

By decree, the crossed product $L_B(\mathcal{H})\rtimes_\varepsilon Sp(8,\C)$  models the abstract quantum $C^\ast$-algebra $\mathfrak{A}_L$ subject to the Born rule. Recall the hypothesis that the nature of $\mathfrak{A}_L$ is made known by way of the Born rule which induces a `topological localization' in the group of units of $\mathfrak{A}_L$. If this notion is going to be physically consistent, then $Sp(8,\C)$ better be contained in the model $C^\ast$-algebra $L_B(\mathcal{H})\rtimes_\varepsilon Sp(8,\C)$:

\begin{theorem}
Let $U(A\rtimes_\varepsilon G)$ denote the unitary group of units in $A\rtimes_\varepsilon G$. Define $i_G:G\rightarrow U(A\rtimes_\varepsilon G)$ by $(i_G(\tilde{g})\mathrm{f})({g})=\varepsilon_{\tilde{g}}(\mathrm{f}(\tilde{g}^{-1}{g}))$. Then $i_G$ is an injective unitary-valued homomorphism such that
\begin{equation}
\pi_{(\Bold{\mu})}\rtimes\rho_{(\Bold{\mu})}(i_G({g}))
=\rho_{(\Bold{\mu})}({g})\,,\,\,\,\,\forall g\in G\;.
\end{equation}
\end{theorem}

\emph{Proof}:
First compute
\begin{eqnarray}
\pi_{(\Bold{\mu})}\rtimes\rho_{(\Bold{\mu})}(i_G(\tilde{g})\mathrm{f})
&=&\int_G\pi_{(\Bold{\mu})}(i_G(\tilde{g})\mathrm{f}(g))\rho_{(\Bold{\mu})}(g)\;d\mu(g)\notag\\
&=&\int_G\pi_{(\Bold{\mu})}(\varepsilon_{\tilde{g}}(\mathrm{f}(\tilde{g}^{-1}{g})))\rho_{(\Bold{\mu})}(g)
\;d\mu(g)\notag\\
&\stackrel{g\rightarrow \tilde{g}g}{=}&\int_G\pi_{(\Bold{\mu})}(\varepsilon_{\tilde{g}}(\mathrm{f}(g)))\rho_{(\Bold{\mu})}(\tilde{g}g)
\;d\mu(g)\notag\\
&=&\int_G\rho_{(\Bold{\mu})}(\tilde{g})\pi_{(\Bold{\mu})}(\mathrm{f}(g)))
\rho_{(\Bold{\mu})}(\tilde{g}^\ast)\rho_{(\Bold{\mu})}(\tilde{g}g)
\;d\mu(g)\notag\\
&=&\rho_{(\Bold{\mu})}(\tilde{g})\cdot\pi_{(\Bold{\mu})}\rtimes\rho_{(\Bold{\mu})}(\mathrm{f})\;.
\end{eqnarray}
This implies $i_G({g_1g_2})=i_G({g_1})i_G({g_2})$ and $i_G({g^{-1}})=i_G({g})^{-1}$. Moreover, from the definition of the norm, $\|i_G({g})\mathrm{f}\|=\|\mathrm{f}\|$. So $i_G$ is a homomorphism and $i_G({g})$ extends to all of $\pi_{(\Bold{\mu})}\rtimes\rho_{(\Bold{\mu})}$.

Now use the definitions of the $\ast$-product and involution to compute
\begin{eqnarray}
((i_G(\tilde{g})\mathrm{f}_1)^\ast\ast\mathrm{f}_2)(g')
&=&\int_G\varepsilon_{g\tilde{g}}(\mathrm{f}_1(\tilde{g}^{-1}g^{-1})^\ast)\Delta(g^{-1})
\varepsilon_g(\mathrm{f}_2(g^{-1}g'))\;d\mu(g)\notag\\
&\stackrel{g\rightarrow g\tilde{g}^{-1}}{=}&
\int_G\varepsilon_{g}(\mathrm{f}_1(g^{-1})^\ast)\Delta(g^{-1})
\varepsilon_{g\tilde{g}^{-1}}(\mathrm{f}_2(\tilde{g}g^{-1}g'))\;d\mu(g)\notag\\
&=&\int_G\mathrm{f}^\ast_1(g)
\varepsilon_{g}(i_G(\tilde{g}^{-1})(\mathrm{f}_2(\tilde{g}g^{-1}g'))\;d\mu(g)\notag\\
&=&(\mathrm{f}^\ast_1\ast i_G(\tilde{g})^{-1}\mathrm{f}_2)(g')\;.
\end{eqnarray}
On the other hand, $(i_G(\tilde{g})\mathrm{f}_1)^\ast\ast\mathrm{f}_2=\mathrm{f}_1^\ast\ast(i_G(\tilde{g})^\ast\mathrm{f}_2)$ so $i_G(g)$ is unitary and $i_G$ maps to a (sub)group of units in $A\rtimes_\varepsilon G$. (One can go further and show that $i_G$ is strictly continuous.\cite[pg. 54]{W})
$\QED$

\section{The appearance of geometry}\label{expected geometry}
The crossed product elements associated with points in the coset space $Z$ correspond to observables $O_Z\in L_B(\mathcal{H})\rtimes_\varepsilon Sp(8,\C)$. It is clear that the spectra $\mathcal{Z}:=\sigma(O_Z)$ of these mutually commuting observables can be thought of as a configuration space characterizing the CS parameter space. Likewise, the spectra $\sigma(O_Z)\times\sigma(O_{P^\C/U(4)})$ can be viewed as an associated complex cotangent bundle $T^\ast\mathcal{Z}$ with $\mathrm{dim}_\C(\mathcal{Z})=10$; assuming the spectra form a topological space. However, the geometry we wish to expose is associated with a real subspace of $T^\ast\mathcal{Z}$.

To that end, note that $Sp(8,\C)$ contains an inner, anti-involutive automorphism (equivalently an almost complex structure) $j$. This almost complex structure obviously transfers to $\mathcal{V}_{(\mu)}$ via $\varrho$. Consequently, for any complex $\mathcal{V}^\C_{(\mu)}$, there exists a basis that diagonalizes $J:=\varrho({j})$ and induces the decomposition $\mathcal{V}^\C_{(\mu)}=\mathcal{V}_{(\mu)}^{+}\oplus\mathcal{V}_{(\mu)}^{-}$ where
\begin{equation}
\mathcal{V}_{(\mu)}^{\pm}:=\left\{\Bold{v}\in\mathcal{V}^\C_{(\mu)}\,|\,J\Bold{v}=\pm i\Bold{v}\right\}
\;,\;\;\;\;\forall\Bold{v}\in\mathcal{V}^\C_{(\mu)}\;.
\end{equation}
Hence, ${j}$ provides a means to transfer objects formulated in the context of $Sp(8,\C)$ into objects relevant to $Sp(8,\R)$ and vice versa. Evidently, the relevant CS spawned by a choice of $j$ will be a sub-representation parametrized by a real sub-space $X\subset Z$ with $\mathrm{dim}_\R(X)=10$. Via the crossed product, this leads to associated operators  whose spectra, under suitable conditions, can be interpreted as a real cotangent bundle $T^\ast\mathcal{X}\subset T^\ast\mathcal{Z}$, and so we need to understand how $Sp(8,\R)$ can be realized from $Sp(8,\C)$.

Recall that $\mathfrak{Sp}(8)$ is endowed with a non-degenerate, bi-linear, symmetric form $B$ --- the Cartan-Killing metric. Together with $j$, this defines a symplectic form by $\mathit{\Omega}(\cdot,\cdot):=B(\cdot,{Ad}(j)\cdot)$. The metric and symplectic structures can be combined to construct a sesquilinear form $h:\mathfrak{Sp}(8,\C)\times\mathfrak{Sp}(8,\C)\rightarrow\C$ by
\begin{equation}
h(\mathfrak{g}_1,\mathfrak{g}_2)=B(\mathfrak{g}_1,\mathfrak{g}_2)-i\mathit{\Omega}(\mathfrak{g}_1,\mathfrak{g}_2)
\,,\;\;\;\;\;\forall \mathfrak{g}_1,\mathfrak{g}_2\in\mathfrak{Sp}(8).
\end{equation}
This sesquilinear form $h$ gives rise to multiple real-valued forms that characterize the subgroup $Sp(8,\R)$; to which we now turn.

The group manifold $\M[Sp(8,\R)]$ possesses the simple topology $\M[U(4)]\times\R^{20}$.\cite{AN} It can be covered by five disjoint domains distinguished by the five pseudo-Euclidean real forms induced by $h$. Explicitly, the exponentiated real Cartan subalgebras are isomorphic to $\mathbb{T}^k\times\R^{k'}$ where $k+k'=4$, and they are equipped with metrics having signatures $(k,k')$.\cite{KM}

These considerations, together with the observation that  $e$ and $j$ are the only two \emph{inner} (anti)involutive automorphisms of $Sp\,(8,\C)$,  motivate the notion of pre-geometry:
\begin{definition}\label{STA}
Let $\{U_i,\varphi_i\}$ be a local trivialization of $\mathcal{W}_{\mathcal{V}}$.
Pre-geometry $\mathcal{G}\subset L(\mathcal{H})$ is
generated by the image under $\rho'_\R$ of $\mathfrak{e}^\dag_{ab}$ (identified with the associated ten left-invariant vector fields
on $U_i$) together with their inner (anti)involutive automorphisms, that is
\begin{equation}
\mathcal{G}:=\mathrm{span}_\R\left\{E,E_{a}, E_{a,b},
E_{-a},J\right\},\;\;\;\;a\neq b\in\{1,2,3,4\};
\end{equation}
where $E:=\rho_\R(e)=Id$, $E_{a}:=\rho'_\R(\mathfrak{e}^\dag_a)$,
$E_{a,b}:=\rho'_\R(\mathfrak{e}^\dag_{a,b})$, and
$J:=\rho_\R(j)$.
\end{definition}

Use the pre-geometry to define $\mathit{\Pi}_a:=\rho'_\R(\mathfrak{e}^\dag_a-\mathfrak{e}_a)$ and $\mathit{\Pi}_{a,b}:=\rho'_\R(\mathfrak{e}^\dag_{a,b}-\mathfrak{e}_{a,b})$. Combine them into $\mathit{\Pi}_i$ with $i\in\{1,\ldots,10\}$. Then the CS ground-state inner product in each fiber $\mathcal{W}_z$ (\textit{z}GIP) given by
\begin{equation}
\frac{(\psi_0|\pi'_z\{\mathit{\Pi}_{i},\mathit{\Pi}_{j}^\dag\}\psi_0)_{\mathcal{W}_z}}
{(\psi_0|\psi_0)_{\mathcal{W}_z}}
=:(g_{ij})_z
\end{equation}
defines a real symmetric form where $\psi_0$ is a ground state and $(\cdot|\cdot)_{\mathcal{W}_z}$ is the Hermitian inner product in $\mathcal{W}_z$. The definition is valid for ground states in each $\mathcal{V}^{(r)}_{(\mu)}$, and since $\mathcal{W}_{(\Bold{\mu})}$ is a direct sum of $\mathcal{V}^{(r)}_{(\mu)}$ the definitions hold for a ground state $\Bold{v}_{\Bold{w}_-}\in\mathcal{W}_{(\Bold{\mu})}$ as well. Note that $\mathrm{tr}(\{\mathit{\Pi}_{a},\mathit{\Pi}_{a}^\dag\})\propto \mathrm{tr}(\rho'_\R(\mathfrak{h}_a^2))$ has signature $(k,k')$.\footnote{Aside from $(k,k')$ and $(k',k)$ being physically equivalent, there is nothing (in the kinematics at least) that distinguishes one signature from another. In this regard, we note that amplitude analyses in gauge QFTs profitably employ space-time signatures  of $(4,0)$ and $(2,2)$ for massive and massless gauge bosons and $(3,1)$ for matter fields.}

Momentarily ignoring that we are not developing dynamics in this paper, a non-trivial evolution of the ground state $|\psi_{U(t)}\rangle:=|U(t)\psi_{0}\rangle$ will induce a change
\begin{equation}
(g_{ij}^{U(t)})_z
:=\frac{(\psi_0|U^{-1}(t)\pi'_z\{\mathit{\Pi}_{i},\mathit{\Pi}_{j}^\dag\}\,U(t)\psi_0)_{\mathcal{W}_z}}
{(\psi_0|\psi_0)_{\mathcal{W}_z}}
=\frac{(\psi_{U(t)}|\pi'_z\{\mathit{\Pi}_{i},\mathit{\Pi}_{j}^\dag\}\psi_{U(t)})_{\mathcal{W}_z}}
{(\psi_0|\psi_0)_{\mathcal{W}_z}}\neq(g_{ij})_z\,;
\end{equation}
so this symmetric form is dynamical unless the ground state coincides with the vacuum. Similarly, non-trivial dynamics induces a time-dependent anti-symmetric form
\begin{equation}
({\mathit{\Omega}_{ij}^{U(t)}})_z
:=\frac{(\psi_{U(t)}|\pi'_z[\mathit{\Pi}_{i,j},\mathit{\Pi}_{i,j}^\dag]\psi_{U(t)})_{\mathcal{W}_z}}
{(\psi_0|\psi_0)_{\mathcal{W}_z}}
\end{equation}
and a time-dependent almost complex structure
\begin{equation}
({J}^{U(t)})_z
:=\frac{(\psi_{U(t)}|\pi_z(J)\psi_{U(t)})_{\mathcal{W}_z}}
{(\psi_0|\psi_0)_{\mathcal{W}_z}}\;.
\end{equation}
Notice that both forms are quartic in the creation/annihilation operators coming from the bosonic Fock space realization of $\mathfrak{Sp}(8)$. For simplicity, henceforth assume they are non-degenerate.

\begin{theorem}
Let $U(t)\in U(L_B(\mathcal{H})\rtimes_\varepsilon Sp(8,\C))$ be a time-dependent evolution operator in the unitary group of units in $L_B(\mathcal{H})\rtimes_\varepsilon Sp(8,\C)$. Suppose $\rho'(\mathfrak{z}_+(t))\in L(\mathcal{H})$ for all $\mathfrak{z}_+(t)=Ad(U(t))\mathfrak{z}_+$. The geometric objects $(g_{ij}^{U(t)})$, $({\mathit{\Omega}_{ij}^{U(t)}})$, and $({J}^{U(t)})$ as defined above are dynamical if the reference CS ground state $\psi_0$ is not the vacuum.
\end{theorem}

\emph{Proof}:
First note that these expectations can only depend parametrically on the spectra $\mathcal{Z}$ because $\rho'(\mathfrak{p})\psi_0$ only transforms the $|\Bold{\mu})$ component of $\Bold{\psi}_0(z)$. Indeed, $\rho'(\mathfrak{e}^\dag_{a,b})$ annihilates the ground state, $\rho'(\{\mathfrak{u}_{ab}\})$ is unitary, and $\rho'([\mathfrak{e}_a,\mathfrak{e}^\dag_b])\propto\rho'(\delta_{ab}\mathfrak{h}_a)$ is normal. Hence, $Ad(p)\rho'([\mathfrak{e}_a,\mathfrak{e}^\dag_b])$ can be diagonalized by a unitary similarity transformation on $\mathcal{W}_{(\Bold{\mu})}$, which implies that the adjoint action of the subgroup $P^\C$ just corresponds to a change of coordinate basis in $\mathcal{W}_{(\Bold{\mu})}$.

Thesis 2.2 dictates the time-evolution $\mathfrak{z}_+(t)=Ad(U(t))\mathfrak{z}_+$ for $\mathfrak{z}_+\in\mathfrak{Z}_+$. Note that $\mathfrak{z}_+(t)$ is in general an \emph{unbounded} operator. However, $\rho'(\mathfrak{z}_+(t))\in L(\mathcal{H})$ by assumption which implies it is adjointable on a suitable domain in $\mathcal{H}$ and we can extract its self-adjoint piece, say $\mathfrak{x}(t)=\mathfrak{x}(t)^\ast$. The spectra of all  $\mathfrak{x}(t)$  are real and time-dependent in general. Finally, from the definition of the CS vacuum and ground state, it follows that $(z;\Bold{\mu}|U(t)\varphi_0\rangle=|\varphi_0\rangle$ but $(z;\Bold{\mu}|U(t)\psi_0\rangle=|\psi_{U(t)}\rangle$ is time-dependent in general precisely because $U(4,\C)$ is non-abelian.
$\QED$

Evidently, for dynamical $Sp(8,\R)$, the \textit{z}GIPs of  $\mathcal{G}$ characterize the geometry of an evolution-dependent cotangent bundle
$T^\ast\mathcal{X}^{U(t)}:=\rho(Ad(U(t))T^\ast\mathcal{X}$ that can be interpreted as a phase-space state in the sense of the GNS construction, assuming it remains a topological space under evolution. For want of a better name, we will call it the `apparent phase space'.\footnote{In the limit of large systems such that $N=\mathrm{dim}_\C(\mathcal{W}_{(\Bold{\mu})})\rightarrow\infty$, `apparent phase space' is posited to become classical, but a general phase-space state would certainly not look classical for all systems. Note that $g$ and $\mathit{\Omega}$ are not related through ${J}$ so the expected
geometry is not K\"{a}hler in general.}

In general, CS are parametrized by dynamical symmetric $4\times 4$ matrices where expectations of $E_{a,b}$ can be interpreted as directed-area elements: there is no need to hide the geometry induced by these degrees of freedom.\footnote{Indeed, if we were considering dynamics here the off-diagonal degrees of freedom would be anticipated to be relevant to rotational dynamics and to generate $4$-d volume elements through suitable combinations of directed-area/directed-area interactions.}. But, for an evolving system that stays near the ground state, one might anticipate the influence of $\mathfrak{e}_{a,b}$ on phase-space geometry to be small if substantial \textit{z}GIPs of $E_{a,b}$ require large quantum numbers or sizable tensored representations. Regardless the reason, let us suppose the evolving ground state has support only along the diagonal of $\Bold{Z}$. Specifically, go to the matrix picture and assume $\mathcal{Z}=\sigma(O_{\Bold{Z}})$ is a topological space. Let $\mathbb{M}^\C\subset \mathcal{Z}$ denote the $\mathrm{dim}_\C(\mathbb{M}^\C)=4$ sub-space associated with diagonal elements in $\Bold{Z}$. Suppose $\mathbb{M}\subset \mathbb{M}^\C$ is a real  sub-space defined with the help of $(J^{U(t)})$. Then $\mathbb{M}$ and $g_{ab}(m):=(g_{ij})_z|_{T\mathbb{M}}$
model a $\mathrm{dim}_\R(\mathbb{M})=4$, signature $(k,k')$ pseudo-Riemannian manifold that represents the \textit{z}GIPs of the pre-geometry with respect to this particular CS ground state.

As described, geometry ``appears'' by virtue of CS inner products and so is intimately tied to the dynamics of the reference ground state. Note that similar notions and constructions can be applied to different starting groups which also lead to apparent geometry, e.g. the Heisenberg group or Poincar\'{e}. The distinguishing features of $Sp(8,\R)$, compared to Heisenberg or Poincar\'{e}, are the $10$-d configuration space with diagonal metric of signature $(k,k')$ and the non-abelian subgroup $U(4,\C)$.

\section{Discussion}
To present a concise picture, we have not mentioned several important aspects of the quantum kinematics of $Sp(8,\C)$ --- not to mention the dynamics. In summary, it seems appropriate to list some features of the full quantum mechanics not directly addressed in the body of the paper.
\begin{itemize}
\item Quantization via the crossed product $C^\ast$-algebra based on $Sp(8,\C)$ is (in our opinion) conceptually cleaner than quantizing an assumed classical system. For one thing, operators do not depend on configuration space so the associated unpleasant consequences that haunt QFT go away. Notably, although the structure of this relatively elementary approach looks to be quite limited, it  spawns a broad array of promising implications.

\item The crossed product elements associated with $Sp(8,\C)/U(4,\C)$ can be interpreted as a non-commutative phase space. Importantly, this phase space possesses a $10$-d \emph{commutative} real subspace with associated spectrum $\mathcal{X}\subset\mathcal{Z}$. For evolving ground states that stay ``near'' the vacuum and only have non-trivial support on the diagonal of $\Bold{Z}$, the emergence of $ISO(k,k')$ symmetry is a consequence of  an \emph{emergent} (since it is now induced by dynamics) $\R^{k,k'}$ configuration space. For such states in the $(k,k')\equiv (3,1)$ domain, spinor/vector matter and causality dynamically emerge along with the Minkowski space.

\item Since the apparent geometry associated with the signature $(3,1)$ encodes a ``time'' coordinate, this framework essentially includes two logically distinct notions of time (imaginary parameters); an ``evolution-time'' which is shared by observers governed by the family $G_\Lambda$, and an ``apparent-time'' that can be measured/observed in each closed system governed by $G_\lambda$.  Insofar as the spectrum of the operator that yields ``apparent-time'' is continuous and the ``evolution-time'' of the evolution operator can be reparametrized, it seems an observer governed by a particular $G_\lambda$ cannot distinguish (by measurement) between the two distinct ``times''.

\item $U(4,\C)$ contains the gauge group of the Standard Model as a subgroup. If the quantum mechanics of $Sp(8,\C)$ is at all realistic, then $U(4,\C)$ offers physics beyond the Standard Model.

\item Ostensibly one could construct various classical $\rightarrow$ quantum effective QFTs on $\mathcal{X}$ that would represent particular aspects of the complete quantum theory. In such effective QFTs, the momentum-type and internal-type degrees of freedom would be artificially segregated. But in the full quantum theory they are interchangeable --- thus blurring the line between momentum and internal quantum numbers. Indeed, if we are allowed to identify elementary `particles' with the \emph{basic} dominant-integral lowest-weight representations, then there is no distinction between matter and force at the quantum level.

\item The apparent geometry associated with $\mathcal{X}$ is dynamical, but that doesn't necessarily mean the geometry is consistent with classical General Relativity. However, there are indications that it might be.\cite[appx. A]{LA5}

\item Dynamics in the Heisenberg picture leads to a quantum matrix model. Similar models\footnote{The difference to emphasize is that the  CS-representation matrices are comprised of operators coming from the crossed product while the matrices in M(atrix) models are comprised of complex parameters and must be subsequently quantized.} are known to encode remarkable complexity. Superficially at least, there are obvious coincidences and perhaps some tangential contact with M(atrix)-theory.

\item The classical approximation appears to describe physics in ten dimensions; six of which correspond to directed area elements. In particular, the classical Boltzmann equation becomes a matrix Boltzmann equation that would presumably offer a superior model of (semi)classical physics --- especially in the presence of vortex dynamics --- if quantum $Sp(8,\C)$ is viable.
\item $Sp(8,\R)$ and $SO(9,\R)$ are Langlands dual and we suspect that a complete treatment should include both groups and perhaps even their supersymmetric parent group $OSp(9,8)$.
\end{itemize}

Before closing, it is worthwhile to return to the introduction. Since we start with $G_\Lambda=Sp(8,\C)$ which already has a manifold structure, isn't the purported direct quantization really just classical $\rightarrow$ quantum in disguise? Aren't we just quantizing a group manifold? The answer is no. The key difference is that the quantum dynamics are a consequence of $i_G(Sp(8,\C))$ being contained in the crossed product, and not the result of a classical model of dynamics (or functions) on $Sp(8,\C)$ being promoted to a non-commutative algebra. This is crucial, since then $Sp(8,\C)$ controls both kinematics and dynamics, and there is no underlying classical manifold that must be consistent with the quantum $\rightarrow$ classical correspondence. It is precisely these two points where emergent geometry/gravity models seem to become physically ambiguous or unrealistic \cite{CAR}.

In retrospect, it is clear why geometry appears (as opposed to emerges). Its roots were there all the time: Evidence of $Sp(8,\C)$ is buried in $\mathcal{A}_L$, and the geometry of $\mathbb{M}[Sp(8,\C)]$ is a reflection of the presumed Lie bracket structure in $\mathcal{A}_L$. Still, geometry as we understand it only appears through expectations with respect to \emph{induced} representations; and evolution of the apparent geometry, being governed by inner automorphisms, is automatically consistent with both kinematics and dynamics.\footnote{There is a \emph{narrow} analogy between induced/discrete-series representations of $Sp(8,\R)$ and Ads/CFT duality. Being UIRs, the infinite-dimensional discrete-series representations can be interpreted as particle excitations in $0$-d. Meanwhile, the CS induced representations can be interpreted as state-vector fields in $10$-d or even as state-vector fields over symmetric $4\times 4$ matrices. Evidently, being equivalent descriptions, some of the group information contained in the infinite collection of quantum numbers of the discrete-series representation gets transferred to a dynamical configuration space in the induced representation; where the dynamics reflect the \emph{non-compact} group relationships originally encoded in the discrete quantum numbers. These two alternative descriptions are not a {duality} in the strict sense of Ads/CFT, but their equivalence does uphold the notion that sometimes an infinite-dimensional collection of quantum excitations can be understood and interpreted in terms of very different configuration spaces. Maybe a classical $\rightarrow$ quantum re-creation of the semi-classical approximation of the two representations would exhibit a gauge/gravity duality, and the web of dualities discovered in recent decades reflects ambiguities inherent in classical $\rightarrow$ quantum models.}

\end{document}